\documentclass[aps, pre, singlecolumn, eqsecnum, a4paper]{revtex4}

\usepackage{epsfig}
\usepackage{amssymb}
\usepackage{amsmath}
\usepackage{bm}
\usepackage{color}
\usepackage[colorlinks,bookmarks=false,citecolor=blue,linkcolor=red,urlcolor=blue]{hyperref}

\newcommand{\be}{\begin{equation}}
\newcommand{\ee}{\end{equation}}

\begin{document}

\title{Validity of the GGE for quantum quenches from interacting to noninteracting models}

\author{Spyros Sotiriadis, Pasquale Calabrese}

\affiliation{Dipartimento di Fisica dell'Universit\`a di Pisa and INFN, Pisa, Italy}

\begin{abstract}
In the majority of the analytical verifications of the conjecture that the Generalised Gibbs Ensemble
describes the large time asymptotics of local observables 
in quantum quench problems, both the post-quench
and the pre-quench Hamiltonians are essentially noninteracting. We
test this conjecture studying the field correlations in the more general
case of an arbitrary pre-quench Hamiltonian, while keeping the post-quench
one noninteracting. We first show that in the previously studied special
case of a noninteracting pre-quench Hamiltonian, the validity of the
conjecture is a consequence of Wick's theorem. 
We then show that it is also valid in
the general case of an arbitrary interacting pre-quench Hamiltonian, 
but this time as a consequence of the cluster decomposition property of the initial state, which is a fundamental principle
for generic physical states. 
For arbitrary initial states that do not satisfy the cluster decomposition property, 
the above conjecture is not generally true. 
As a byproduct of our investigation we obtain an analytical
derivation of earlier numerical results for the large time evolution
of correlations after a quantum quench of the interaction in the Lieb-Liniger
model from a nonzero value to zero. 
\end{abstract}
\maketitle

\section{Introduction}

One of the fundamental principles of statistical mechanics is that a generic isolated \emph{classical} system in the thermodynamic limit, 
prepared in an arbitrary initial state, would evolve so as to maximise its entropy \cite{Jaynes1,Jaynes2}, 
that is for large times it would tend to thermal equilibrium described by the microcanonical ensemble,
with a total energy equal 
to the initial one. An obvious exception is provided by the integrable systems, i.e.
systems that possess a set of integrals of motion equal in number to their degrees of freedom. 
Maximisation of the entropy under all the additional conservation constraints,
would lead to a generalised  thermal equilibrium, in which all other constraints are satisfied too.

But what happens in a quantum system? In an isolated system the time evolution is unitary, so
if the system is initially prepared in a pure state it will always remain in a pure state, instead of a statistical ensemble. 
In fact, the unitary evolution is periodic or quasiperiodic, that is, after
sufficiently large time, the system will return to its initial state
or arbitrarily close to it. However, subsystems of the whole system
are not isolated and therefore are described by a reduced density matrix
that is not pure and may well be equivalent to a statistical ensemble.
The same is true for local physical observables, whose expectation
values are given by traces over such reduced density matrices. On
the other hand, the period of quantum recurrences typically diverges
with the system size, so that considering first the thermodynamic
and then the large time limit in a suitable well-defined manner (see e.g. \cite{CEF-ii}), it
is possible that the system exhibit stationary behaviour, always at
the level of its subsystems and local physical observables rather
than the whole system or global quantities. Under these clarifications,
the question whether an isolated quantum system thermalises for large
times, when starting from an arbitrary initial state, is a sensible one which has been considered already 
in a series of theoretical and experimental 
works \cite{nonint,kww-06,Rigol08,BKL-10,tetal-11,CCR-11,bdkm-12,sks-13,ekmr-14,r-14,wrdk-14} (see also \cite{revq} for a review).

This question is hard to answer in full generality, since the time
evolution of a general quantum system cannot be calculated exactly.
For one-dimensional integrable quantum systems on the other hand,
this is possible at least in principle. 
In this case it has been proposed \cite{Rigol07},
in analogy to the classical case, that, provided the system tends
to equilibrium in the above sense, its stationary behaviour is described
by a Generalised Gibbs Ensemble (GGE)
in which all integrals of motion have been taken into account and
not only the energy. More explicitly, the GGE is described by a density
matrix 
\be \rho_{\text{GGE}}\sim\exp\Big(-\sum_{j}\lambda_{j}I_{j}\Big),
\ee
in which a separate Lagrange multiplier $\lambda_{j}$ is introduced
for each of the integrals of motion $I_{j}$, in the same way that
in the canonical ensemble the temperature is the Lagrange multiplier
associated to the constraint of conservation of the energy. The values
of $\lambda_{j}$ are fixed by the condition that the ensemble expectation
values of the integrals of motion are equal to their initial values.
The conjecture then is that the stationary expectation values of any
\emph{local} physical observable are equal to their ensemble expectation
values or, equivalently, that the reduced density matrix of any finite
subsystem is equal to the corresponding reduce density matrix of $\rho_{\text{GGE}}$. 

However, this conjecture as stated above contains a subtle ambiguity: which are the integrals of motion
that should be included in $\rho_{\text{GGE}}$? Unlike in classical
integrable systems, where the requirement of existence of a set of
integrals of motion equal in number to the degrees of freedom is a
sufficient definition of integrability, in quantum systems there always
exists an infinite set of integrals of motion which are the projections
onto each of the eigenstates of the Hamiltonian but this fact is not
sufficient for their exact solvability. Obviously, an ensemble with
a density matrix that incorporates all projections onto eigenstates
as integrals of motion would give 
correctly the stationary values of \emph{all} observables of the system (or
their long time averaged values, if they do not equilibrate) for any
arbitrary initial state, but this is a trivial fact that has no connection
with the economy of the maximum entropy principle. 
Such an ensemble would retain all of the information about the initial state,
rather than information about only a minimal set of integrals of motion. 
Indeed, assuming for simplicity that the energy spectrum is non-degenerate, the above ensemble is 
identical to the so-called diagonal ensemble (where the integrals of motion are the projections onto equal-energy eigenspaces) 
and its entropy is always smaller than the entropy of the GGE \cite{pol-11,spr-11,sr-10,v-e,g-13,ckc-13,kbc-14}. (When there are degeneracies, it is possible that one should include additional integrals of motion in order to correctly capture the stationary behaviour as discussed in some particular cases in \cite{f-14}).
The ambiguity in the set of integrals of motion that should be included 
is resolved when we consider what are 
the special characteristics of quantum integrable systems (see e.g.~\cite{cm-11}). 
The most useful for our purposes definition of quantum integrability is based
on the existence of (an infinite set of) \emph{local} integrals of motion
which is what guarantees their exact solvability, through the factorisation
of their scattering matrix, another characteristic property of quantum
integrability. The notion of locality of the integrals of motion (or
conserved charges) is appealing in the context of the above conjecture,
since if the GGE is supposed to describe the reduced density matrix
of any finite subsystem, viewed as an open system in contact to the
rest of the system which plays the role of a bath, then $\rho_{\text{GGE}}$
must be given in terms of local quantities, defined within the spatial
region of the subsystem \cite{CEF-ii,fe-13}.

\subsection{Quantum Quenches and the role of the initial state}\label{sec:A}

From the tests of the conjecture that the GGE describes the stationary behaviour of integrable systems that have been performed until
now, it turns out that the role of the initial state is crucial to 
its validity. Obviously, if the initial state is chosen to be a finite superposition of eigenstates
of the Hamiltonian under which it is let to evolve, there would be no equilibration, since the evolution would be periodic even in the thermodynamic limit. 
A common protocol for setting the system in a `proper' out-of-equilibrium
state is the so-called quantum quench \cite{cc-06}, i.e. an instantaneous change
in the Hamiltonian of the system so that initially it lies in the
ground state, as typically chosen, of the Hamiltonian before the quench.
The GGE has turned out valid for all quantum quench problems
studied so far \cite{cc-07,fm-10,CEF,CEF-i,CEF-ii,bs-08,scc-09,r-09,r-09a,c-06,caz,f-13,eef-12,se-12,ccss-11,
rs-12,CE-08,CE-10,CSC13a,fe-13,Rigol07,RS13,Mossel,Pozsgay11,ce-13,KCC-13,f-14,ck-14}.
However the vast majority of these studies refer to physical models
that are either noninteracting or interacting but exactly equivalent
to noninteracting ones, both before and after the quench. This is
mainly due to the fact that for integrable models with a non-trivial
scattering matrix, which therefore cannot be mapped into noninteracting
ones, both the derivation of GGE predictions and the study of the
time evolution are technically difficult problems that
have been accomplished only in a small number of cases 
\cite{fm-10,Pozsgay11,gge-new1,ce-13,sfm-12,gge-new2,gge-new3,gge-new4,RS13,gge-new5,CK-12,mc-12b,a-12,stm-13,bdwc-14,fcec-13,pos-13,b-14,cd-14}.
Testing analytically the validity of the GGE in such models remains a crucial and very
challenging open problem which should parallel current numerical efforts in the same direction \cite{fcec-13}.

The verification of validity of the GGE in noninteracting models displays
common characteristics in all cases. The system can be decomposed
into an infinite set of noninteracting modes, whose occupation numbers
$n_{k}$ are non-local conserved charges, which however are always linear combination of the local ones \cite{CEF-ii,fe-13}. 
Under certain conditions, the interference of these modes leads to equilibration
for large times. For example, in the cases we will study below, the post-quench dispersion relation of excitations is gapped and the initial state is translationally invariant, which are sufficient conditions for the equilibration. However, these are not necessary conditions, since on the one hand gapless post-quench Hamiltonians also lead, for different reasons, to equilibration \cite{cc-07} and on the other hand the breaking of translational invariance in the pre-quench or the post-quench Hamiltonian does not necessarily prevent equilibration (the presence of localisation in the post-quench Hamiltonian instead is one of the reasons that prevent equilibration, but the GGE may still be applicable for certain observables \cite{loc}). 
As long as equilibration occurs, the stationary values
of two-point correlations of fields are typically given by some \emph{linear}
combination of the values of the conserved charges $\langle n_{k}\rangle_{0}$.
The GGE prediction for the two-point correlations, on the other hand,
is given by the same linear combination but with the GGE values of
the charges $\langle n_{k}\rangle_{\text{GGE}}$ instead. Since these
are by definition equal to their initial values, the actual stationary
correlations and their GGE predictions are automatically and trivially
equal to each other. Analogously, both the stationary values and the
GGE values of all higher-order correlations ($n$-point functions, but 
in general of any observable) can be expressed in terms of the expectation
values of the charges and of their products in the initial state
and in the GGE respectively. The GGE for a noninteracting system
is a Gaussian ensemble and this means that, by virtue of Wick's theorem, the products of the charges are uncorrelated in the
GGE (i.e. their GGE expectation values of the products equal the product
of the expectation values). The same is also true for the initial
state which, being the ground state of a noninteracting Hamiltonian,
is also Gaussian. Consequently, even for higher-order correlations
the GGE gives always the correct predictions. This argument explains generally
the validity of the GGE for noninteracting systems.

We realise that the above reasoning is crucially based on the Gaussianity
of the initial state. Indeed, Wick's theorem, which was employed in order to show the validity of the GGE, 
is valid if and only if the initial state is Gaussian in terms of the post-quench quasiparticle modes. The crucial role of Wick's theorem (in combination with the double limit: thermodynamic followed by large time limit, or time averaging followed by thermodynamic limit) was pointed out in Ref.~\cite{caz} for various quantum quenches in essentially noninteracting models (quantum Ising, XX spin chain and Luttinger model) starting with a Gaussian initial state. 
Gaussianity of the initial state manifests itself in various different forms, which have been identified as being sufficient conditions for the validity of the GGE: quadratic form of  the initial density matrix in terms of the post-quench quasiparticle modes \cite{caz} or factorisation of the initial expectation values of charge products \cite{KE08}. Both the former and the latter form have been shown to be a sufficient (but not necessary) condition for the validity of the GGE also in the more general case of interacting post-quench Hamiltonians \cite{fm-10,RS13,gge-new4}.

What if however the pre-quench Hamiltonian is genuinely interacting,
so that the initial state is non-Gaussian, while the post-quench Hamiltonian
is kept noninteracting? The present study is concerned with precisely
this question, which was also raised in Ref. \cite{CE-10}. 
To compare the stationary values of the higher-order
correlation functions with their GGE values, we can express the former
as a convolution of the initial correlation functions and expand these
as a cumulant expansion, i.e. in connected and disconnected terms.
We find that, when performing the spatial integration of these correlation
functions, it is only a maximally-disconnected part that survives
in the combined thermodynamic and large time limit. This means that
all stationary correlations of higher-order can be derived from solely
the two-point initial correlations, that is from information contained
solely in the values of the conserved charges. We therefore show that
all stationary correlations are exactly equal to their GGE predictions.
The proof is based on the \emph{cluster decomposition principle}, 
which states that the correlations between two subsets of points separated by a distance 
that tends to infinity become disconnected, i.e. \cite{Weinberg}
\be 
\lim_{R\to\infty}\Big\langle \prod_{i}\phi(x_{i})\prod_{j}\phi(x_{j}+R)\Big\rangle =
\Big\langle \prod_{i}\phi(x_{i})\Big\rangle \Big\langle \prod_{j}\phi(x_{j})\Big\rangle.
\ee
This property is a fundamental requirement for generic physical states such as 
the ground state of any physical Hamiltonian, the thermal density matrix etc. 
Therefore, we conclude that neither Gaussianity nor factorisation
of charge products in the initial state are necessary conditions for
the validity of the GGE. Contrarily, even in the much
more general case of a quantum quench from an arbitrary interacting
pre-quench Hamiltonian, the GGE is still valid, this time
as a consequence of a fundamental property of the initial state correlations. 
However, in the case of an initial state that does not satisfy the cluster decomposition property, 
the stationary expectation values of observables would generally retain memory of the initial correlations between conserved charges beyond
their maximally-disconnected part and the GGE would not apply. 
Our results are consistent with and generalise earlier findings \cite{CE-10} obtained by completely different methods.

We stress that the cluster decomposition property is satisfied by the ground states of any
physical Hamiltonian, including, but not limited to, local Hamiltonians. Indeed even Hamiltonians with 
long range interactions may have ground states that satisfy this condition. 
Furthermore, thermal (mixed) states of physical Hamiltonians also satisfy the same property and 
even many of the eigenstates are expected to do so. In fact, numerical results \cite{hr-12} confirm that, starting from excited states 
of interacting Hamiltonians, several stationary observables are well approximated by the GGE.

We demonstrate these ideas in the context of two prototypical theories:
a relativistic and a non-relativistic bosonic field theory in one
spatial dimension with no interaction after the quench. In the first
model, we keep our exposition as general as possible and show that
our findings are insensitive to the form of the post-quench dispersion
relation (and so insensitive to relativistic invariance too) or other details of the particular quench and we identify
the minimal requirements for equilibration to occur and for the rest
of our arguments to hold. In this way, it is clear that our results
are true for a wide range of physical systems that are equivalent
to systems of coupled harmonic oscillators with arbitrary couplings.
In the second model, as a byproduct of our investigation, we obtain
an analytical derivation of earlier discovered numerical results \cite{Mossel}
for the relaxation of density-density correlations in the case of
free non-relativistic bosons starting from an initial state with pointlike
interactions (quantum quench in the Lieb-Liniger model from positive
to zero interaction). It turns out that the relaxation is a power-law
in time and is related to the decay of the initial four-point correlation
function at large distances, which is governed by the Luttinger Liquid
approximation of the Lieb-Liniger model, allowing us to calculate
the exponent of the power law from the Luttinger parameter $K$.

\section{Relativistic bosonic field theory}\label{RB}

We consider the one-dimensional system of harmonic oscillators described
by the Hamiltonian (in momentum space) 
\be
H=\frac{1}{2}\sum_{k}\left(\tilde{\pi}_{k}\tilde{\pi}_{-k}+\omega_{k}^{2}\tilde{\phi}_{k}\tilde{\phi}_{-k}\right).
\ee
As well-known, such a Hamiltonian may describe a relativistic free
field theory, if $\omega_{k}=\sqrt{k^{2}+m^{2}}$ with particle mass
$m$, but we do not need to specify the exact form of the dispersion
relation for our subsequent study. This Hamiltonian will play the
role of the post-quench Hamiltonian in our problem. Since it is free,
the time-evolved field operator $\phi$ (in the Heisenberg picture)
can be written in momentum space as 
\begin{equation}
\phi(x;t)=\frac{1}{\sqrt{L}}\sum_{k}{e^{ikx}\tilde{\phi}_{k}(t)}=\frac{1}{\sqrt{L}}\sum_{k}{e^{ikx}\;\frac{1}{\sqrt{2\omega_{k}}}(a_{k}e^{-i\omega_{k}t}+a_{-k}^{\dagger}e^{+i\omega_{k}t})},
\end{equation}
and the conjugate momentum $\pi$ as 
\begin{equation}
\pi(x;t)=\frac{1}{\sqrt{L}}\sum_{k}e^{ikx}\tilde{\pi}_{k}(t)=\frac{1}{\sqrt{L}}\sum_{k}e^{ikx}\;(-i)\sqrt{\frac{\omega_{k}}{2}}(a_{k}e^{-i\omega_{k}t}-a_{-k}^{\dagger}e^{+i\omega_{k}t}).
\end{equation}
Here $L$ is the system size and we assume periodic boundary conditions,
so that the momenta are given by $k=2\pi n/L$ with $n$ integer (although this assumption is not essential).
The creation-annihilation operators can therefore be expressed in the following form which will be useful later 
\begin{align}
a_{k} & =\sqrt{\frac{\omega_{k}}{2}}\tilde{\phi}_{k}(0)+\frac{i}{\sqrt{2\omega_{k}}}\tilde{\pi}_{k}(0),
\qquad 
a_{-k}^{\dagger}  =\sqrt{\frac{\omega_{k}}{2}}\tilde{\phi}_{k}(0)-\frac{i}{\sqrt{2\omega_{k}}}\tilde{\pi}_{k}(0).
\label{eq:3}
\end{align}

\subsection{The two-point correlation function}

As we anticipated, the calculation of the two-point function is rather insensitive on the initial state and so the 
derivation parallels the one for a quench between free theories \cite{cc-07,scc-09}.
Explicitly, the equal-time two-point function is 
\begin{align}
 & C_{t}^{(2)}(x,y)\equiv\langle\phi(x;t)\phi(y;t)\rangle=\frac{1}{L}\sum_{k_{1},k_{2}}\frac{1}{\sqrt{2\omega_{k_{1}}}}\frac{1}{\sqrt{2\omega_{k_{2}}}}\; e^{ik_{1}x+ik_{2}y} \\
 & \times\Big[\langle a_{k_{1}}a_{k_{2}}\rangle_{0}\; e^{-i(\omega_{k_{1}}+\omega_{k_{2}})t}+\langle a_{-k_{1}}^{\dagger}a_{k_{2}}\rangle_{0}\; e^{+i(\omega_{k_{1}}-\omega_{k_{2}})t}
 +\langle a_{k_{1}}a_{-k_{2}}^{\dagger}\rangle_{0}\; e^{-i(\omega_{k_{1}}-\omega_{k_{2}})t}+\langle a_{-k_{1}}^{\dagger}a_{-k_{2}}^{\dagger}\rangle_{0}\; e^{+i(\omega_{k_{1}}+\omega_{k_{2}})t}\Big],\nonumber
\end{align}
where the index zero means that the expectation values are calculated
on the initial state. Assuming that the latter is translationally
invariant, the above expectation values are zero unless the momenta
$k_{1},k_{2}$ are equal or opposite.
Using the canonical commutation relation $[a_{k},a_{p}^{\dagger}]=\delta_{k,p}$, 
the non zero initial correlators can be parametrised as follows 
\begin{align}
\langle a_{k}a_{q}\rangle_{0} & =A_{k}\delta_{k,-q},\nonumber \\
\langle a_{-k}^{\dagger}a_{q}\rangle_{0} & =B_{k}\delta_{k,-q},\nonumber\\
\langle a_{k}a_{-q}^{\dagger}\rangle_{0} & =(1+B_{-k})\delta_{k,-q},\nonumber\\
\langle a_{-k}^{\dagger}a_{-q}^{\dagger}\rangle_{0} & =A_{-q}^{*}\delta_{k,-q}.
\end{align}
where $A_{k}$ and $B_{k}$ are functions that depend on the particular
initial state. From the definition of $B_{k}$ we have 
\begin{equation}
B_{-k}=\langle a_{k}^{\dagger}a_{k}\rangle_{0}=\langle n_{k}\rangle_{0},
\end{equation}
which is the momentum occupation number in the initial state.

From all the above, we obtain 
\begin{align}
C_{t}^{(2)}(x,y) & =\frac{1}{L}\sum_{k}\frac{1}{2\omega_{k}}\; e^{ik(x-y)}\Big[\langle a_{k}a_{-k}\rangle_{0}\; e^{-2i\omega_{k}t}+\langle a_{-k}^{\dagger}a_{-k}\rangle_{0}+\langle a_{k}a_{k}^{\dagger}\rangle_{0}+\langle a_{-k}^{\dagger}a_{k}^{\dagger}\rangle_{0}\; e^{+2i\omega_{k}t}\Big].
\end{align}
In the thermodynamic limit $L\to\infty$, the sum in the above expression
becomes an integral over continuous momenta 
\begin{align}
C_{t}^{(2)}(x,y) & =\int\frac{dk}{2\pi}\frac{1}{2\omega_{k}}\; e^{ik(x-y)}\Big[\langle a_{k}a_{-k}\rangle_{0}\; e^{-2i\omega_{k}t}+\langle a_{-k}^{\dagger}a_{-k}\rangle_{0}+\langle a_{k}a_{k}^{\dagger}\rangle_{0}+\langle a_{-k}^{\dagger}a_{k}^{\dagger}\rangle_{0}\; e^{+2i\omega_{k}t}\Big]. \label{Ct}
\end{align}

Finally we take the long time limit $t\to\infty$ of the above expression.
For a massive post-quench dispersion relation, $\omega_{k}=\sqrt{k^{2}+m^{2}}$
with $m\neq0$, the stationary phase method shows that the oscillating
time-dependent terms in the above integral vanish and therefore the two-point
function becomes stationary. Notice that this argument is
not sensitive to the particular quench we consider; it only requires
a \emph{gapped} post-quench dispersion relation with a single smooth
local minimum at $k=0$, but not necessarily relativistic.
In fact even for a gapless dispersion relation, the model will generically equilibrate \cite{cc-07,sc-10}, even though the 
stationary phase argument does not straightforwardly apply. The difference is that in this case equilibration refers to the correlations of the vertex operators which are the real physical observables in the gapless case.

As long as the above condition is satisfied, the two-point function
equilibrates and we will now express its stationary value in terms of $\langle n_{k}\rangle_{0}$. 
Having recognised which terms of (\ref{Ct}) survive in the large time limit, we will keep them in their finite volume
form as sums over discrete momenta, since in the next step we obtain a $\delta(0)$ term which makes sense only
within the finite volume expression. We therefore find
\begin{align}
C_{\infty}^{(2)}(x,y) & = \frac{1}{L}\sum_{k} \frac{1}{2\omega_{k}}e^{ik(x-y)}\ensuremath{\left(1+\langle n_{-k}\rangle_{0}+\langle n_{k}\rangle_{0}\right)}.
\label{eq:2pt_lt}
\end{align}
or returning to the thermodynamic limit
\begin{align}
C_{\infty}^{(2)}(x,y) & =\int\frac{dk}{2\pi}\frac{1}{2\omega_{k}}e^{ik(x-y)}\ensuremath{\left(1+\langle n_{-k}\rangle_{0}+\langle n_{k}\rangle_{0}\right)}.
\end{align}

We will now calculate the GGE prediction for the two-point correlation
function. The GGE is given, always in the thermodynamic limit, by
the density matrix 
\begin{equation}
\rho_{\rm GGE}=Z^{-1}{\exp\ensuremath{\left(-\int\frac{dk}{2\pi}\lambda_{k}n_{k}\right)}},\label{gge}
\end{equation}
where 
\begin{equation}
Z=\text{Tr}\left\{ \exp\ensuremath{\left(-\int\frac{dk}{2\pi}\lambda_{k}n_{k}\right)}\right\}, \label{gge2}
\end{equation}
and the Lagrange multipliers $\lambda_{k}$ are defined through the
requirement that the values of the charges $n_{k}$ in the GGE are
equal to their initial values
\begin{equation}
\langle n_{k}\rangle_{\rm GGE}=\langle n_{k}\rangle_{0}.\label{gge3}
\end{equation}
The GGE value of the two-point function is then
\begin{align}
C_{\text{GGE}}^{(2)}(x,y) & \equiv\langle\phi(x)\phi(y)\rangle_{\text{GGE}}
=\int\frac{dk}{2\pi}\frac{1}{2\omega_{k}}e^{ik(x-y)}\ensuremath{\left(\langle a_{-k}^{\dagger}a_{-k}\rangle_{\text{GGE}}+\langle a_{k}a_{k}^{\dagger}\rangle_{\text{GGE}}\right)}\nonumber \\
 & =\int\frac{dk}{2\pi}\frac{1}{2\omega_{k}}e^{ik(x-y)}\ensuremath{\left(1+\langle n_{-k}\rangle_{\text{GGE}}+\langle n_{k}\rangle_{\text{GGE}}\right)},
 \label{2pt_gge}
\end{align}
since as expressed in (\ref{gge}) the GGE is diagonal in the momentum
basis.

The last result is obviously equal to (\ref{eq:2pt_lt}), by virtue
of the defining condition (\ref{gge3}) of the GGE. Note that as long
as the two-point function equilibrates for long times, its asymptotic
value is automatically given by the GGE prediction with no further
assumption, simply because the only information of the initial state
on which it depends are the values of the charges which are fixed
in the GGE by definition. We also stress that the only assumption required
for the equilibration and therefore also for the verification of the
equality between (\ref{2pt_gge}) and (\ref{eq:2pt_lt}) was that
the post-quench dispersion relation has a gap and a single minimum (note however that, as already mentioned, this condition can be released);
no further information about the initial state is required (we also
assumed the generally applicable property of translational invariance).
By the generally assumed symmetry of the initial state and the evolving
Hamiltonian under parity (coordinate space reflections), we have $\langle n_{-k}\rangle_{0}=\langle n_{k}\rangle_{0}$,
but we did not need to use this fact. We therefore arrive at the general
conclusion that \emph{the stationary expression of the two-point function,
in the case of noninteracting evolution, will always be trivially
described by a GGE with the momentum occupation numbers as conserved
charges, for any initial state} (this statement strictly holds only as long as the single particle spectrum is non-degenerate; in the opposite case particular care should be taken in order to fix the appropriate set of integrals of motion as discussed in \cite{f-14}). 

Before we proceed further, we will derive a direct relation between
the large time asymptotic value of the field correlations and the
initial ones in their coordinate space form. From (\ref{eq:3}) the
operator $n_{k}+n_{-k}$ that appears in (\ref{eq:2pt_lt}) can be
written in terms of the field $\phi$ and its time derivative $\dot{\phi}=\pi$ as
\begin{align}
  n_{k}+n_{-k}&=a_{k}^{\dagger}a_{k}+a_{-k}^{\dagger}a_{-k}
 \nonumber \\& 
 =\frac{1}{2}\left[\omega_{k}\left(\tilde{\phi}_{-k}\tilde{\phi}_{k}+\tilde{\phi}_{k}\tilde{\phi}_{-k}\right)+i\left(\tilde{\phi}_{-k}\tilde{\pi}_{k}+\tilde{\phi}_{k}\tilde{\pi}_{-k}-\tilde{\pi}_{-k}\tilde{\phi}_{k}-\tilde{\pi}_{k}\tilde{\phi}_{-k}\right)+\left(\tilde{\pi}_{-k}\tilde{\pi}_{k}+\tilde{\pi}_{k}\tilde{\pi}_{-k}\right)/\omega_{k}\right]\nonumber \\
 & =\omega_{k}\tilde{\phi}_{k}\tilde{\phi}_{-k}+\tilde{\pi}_{k}\tilde{\pi}_{-k}/\omega_{k}-1,
 \label{eq:1}
\end{align}
where in the last step we used the commutation relations $[\tilde{\phi}_{k},\tilde{\pi}_{q}]=i\delta_{k,-q}$
and $[\tilde{\phi}_{k},\tilde{\phi}_{q}]=[\tilde{\pi}_{k},\tilde{\pi}_{q}]=0$.
Therefore substituting in Eq. (\ref{eq:2pt_lt}) we have 
\begin{align}
C_{\infty}^{(2)}(x,y) & =\frac{1}{2}\int\frac{dk}{2\pi}e^{ik(x-y)}\ensuremath{\left(\left\langle \tilde{\phi}_{k}\tilde{\phi}_{-k}\right\rangle _{0}+\frac{1}{\omega_{k}^{2}}\left\langle \tilde{\pi}_{k}\tilde{\pi}_{-k}\right\rangle _{0}\right)}
=\frac{1}{2}\bigg(C_{0}^{(2)}(x-y)+\intop_{-\infty}^{+\infty}ds\, H(s)D_{0}^{(2)}(x-y-s)\bigg),
\label{eq:Cinfty}
\end{align}
where we defined 
\begin{align}
H(x)\equiv & \int\frac{dk}{2\pi}\frac{e^{ikx}}{\omega_{k}^{2}},\label{eq:H}\\
C_{0}^{(2)}(x)\equiv & C_{0}^{(2)}(0,x),\label{eq:C}
\end{align}
and 
\begin{align}
D_{0}^{(2)}(x) & \equiv\left\langle \pi(0,0)\pi(x,0)\right\rangle =\left.\frac{\partial}{\partial t_{1}}\frac{\partial}{\partial t_{2}}\left\langle \phi(0,t_{1})\phi(x,t_{2})\right\rangle \right|_{t_{1}=t_{2}=0}.
\end{align}
These formulas will be useful later in the comparison of the stationary
four-point function with its GGE prediction.

\subsection{The four-point correlation function}

We now proceed to the calculation of the four-point function which,
according to the discussion in \ref{sec:A}, is the first non-trivial test of the 
conjecture that the GGE describes the stationary behaviour after a quantum quench.
Our main objective is to check \emph{whether the large time four-point
function retains such memory of the initial four-point function that cannot
be derived from the initial two-point function}. If this is true,  the
large time four-point function will not be  described by the GGE (\ref{gge}) because in the latter, Wick's theorem is valid
and therefore the predicted four-point function depends solely on the
two-point function (or in other words on $\langle n_{k}\rangle_{0}$)
and not on any additional information about the initial state. 
The calculation below is largely inspired by one for a very specific case (the 
quench of a Bose gas from zero to infinite interaction \cite{KCC-13}), but as we shall see, 
the reason for its general validity is a deeper physical requirement on the initial state, i.e. the cluster decomposition principle.

The equal time four-point function is 
\begin{align}
  C_{t}^{(4)}(x_{1},x_{2},x_{3},x_{4})&\equiv\langle\phi(x_{1};t)\phi(x_{2};t)\phi(x_{3};t)\phi(x_{4};t)\rangle=
 \nonumber \\& =
 \frac{1}{L^{2}}\sum_{k_{1},k_{2},k_{3},k_{4}}\frac{1}{4\sqrt{\prod_{i=1}^{4}\omega_{k_{i}}}}\; e^{i\sum_{i=1}^{4}k_{i}x_{i}}\sum_{{\text{all }\{\sigma_{i}\}\atop \sigma_{i}=\pm}}\left\langle \prod_{i=1}^{4}a_{-\sigma_{i}k_{i}}^{(\sigma_{i})}\right\rangle _{0}\; e^{i\sum_{i=1}^{4}\sigma_{i}\omega_{k_{i}}t},\label{4pt_1}
\end{align}
where we used the compact notation $a_{k}^{(+)}\equiv a_{k}^{\dagger}$
and $a_{k}^{(-)}\equiv a_{k}$. Taking first the thermodynamic limit,
in which the momentum sums become integrals, and then the large time
limit, we observe that, provided that the stationary phase argument
applies as above for a gapped post-quench Hamiltonian, only terms with no oscillating phase survive, i.e.
those satisfying the condition $\sum_{i=1}^{4}\sigma_{i}\omega_{k_{i}}=0$
with $\sigma_{i}=\pm1$. The latter is satisfied only by terms with
equal number of $a$ and $a^{\dagger}$ operators (which are $4!/(2! 2!)=6$ in number,
out of $2^{4}=16$) and more specifically those for which
the momenta of the $a$ operators match with those of the $a^{\dagger}$
operators in pairs of equal or opposite values. This is because non-polynomial
equations of the form $\sum_{i=1}^{4}\sigma_{i}\omega_{k_{i}}=0$
have only sparse solutions for discretised momenta $k=2\pi n/L$,
that result in negligible contributions in the thermodynamic limit,
unless $\sum_{i=1}^{4}\sigma_{i}=0$ in which case the trivial solutions
in which the $k_{i}$'s appear in pairs of equal or opposite values
give finite (non-vanishing) contributions in the thermodynamic limit. Obviously, in the above we make use of the fact that the dispersion relation is an even function, i.e. $\omega_{-k}=\omega_{k}$. 
For example, the equation $\omega_{k_{1}}+\omega_{k_{2}}=\omega_{k_{3}}+\omega_{k_{4}}$
is satisfied by the terms that contain the expectations values $\langle a_{k_{1}}a_{k_{2}}a_{-k_{3}}^{\dagger}a_{-k_{4}}^{\dagger}\rangle_{0}$
and $\langle a_{-k_{1}}^{\dagger}a_{-k_{2}}^{\dagger}a_{k_{3}}a_{k_{4}}\rangle_{0}$
under the condition that $k_{1}=\pm k_{3}$ and $k_{2}=\pm k_{4}$
or $k_{1}=\pm k_{4}$ and $k_{2}=\pm k_{3}$, while the equation $\omega_{k_{1}}+\omega_{k_{2}}+\omega_{k_{3}}=\omega_{k_{4}}$
does not have any such simple solutions and the corresponding terms
give vanishing contributions in the thermodynamic and large time limit.
Note that we should take first the thermodynamic and then the large
time limit, since in order to apply the stationary phase method, the
sums should have been first written as integrals. However, as in the case of the two-point function, now that we have recognised
which types of terms survive in the large time limit, we will keep them in their finite volume
form, because in the subsequent algebra
some contractions result in $\delta^{2}$ terms that make sense only
within the finite volume expressions.

According to the above we have
\begin{align}
 & C_{\infty}^{(4)}(x_{1},x_{2},x_{3},x_{4})=\frac{1}{L^{2}}\sum_{k_{1},k_{2},k_{3},k_{4}}\frac{1}{4\sqrt{\prod_{i=1}^{4}\omega_{k_{i}}}}\; e^{i\sum_{i=1}^{4}k_{i}x_{i}}
 \times\nonumber \\& 
 \Bigg\{\; \left(\ensuremath{\langle a_{k_{1}}a_{k_{2}}a_{-k_{3}}^{\dagger}a_{-k_{4}}^{\dagger}\rangle_{0}+\langle a_{-k_{1}}^{\dagger}a_{-k_{2}}^{\dagger}a_{k_{3}}a_{k_{4}}\rangle_{0}}\right)
 \Big[\ensuremath{\left(\delta_{k_{1},k_{3}}+\delta_{k_{1},-k_{3}}\right)}\ensuremath{\left(\delta_{k_{2},k_{4}}+\delta_{k_{2},-k_{4}}\right)}+\ensuremath{\left(\delta_{k_{1},k_{4}}+\delta_{k_{1},-k_{4}}\right)}\ensuremath{\left(\delta_{k_{2},k_{3}}+\delta_{k_{2},-k_{3}}\right)}\Big]\nonumber \\
 & +\left(\ensuremath{\langle a_{k_{1}}a_{-k_{2}}^{\dagger}a_{k_{3}}a_{-k_{4}}^{\dagger}\rangle_{0}+\langle a_{-k_{1}}^{\dagger}a_{k_{2}}a_{-k_{3}}^{\dagger}a_{k_{4}}\rangle_{0}}\right)
 \Big[\ensuremath{\left(\delta_{k_{1},k_{2}}+\delta_{k_{1},-k_{2}}\right)}\ensuremath{\left(\delta_{k_{3},k_{4}}+\delta_{k_{3},-k_{4}}\right)}+\ensuremath{\left(\delta_{k_{1},k_{4}}+\delta_{k_{1},-k_{4}}\right)}\ensuremath{\left(\delta_{k_{2},k_{3}}+\delta_{k_{2},-k_{3}}\right)}\Big]\nonumber \\
 & +\left(\ensuremath{\langle a_{k_{1}}a_{-k_{2}}^{\dagger}a_{-k_{3}}^{\dagger}a_{k_{4}}\rangle_{0}+\langle a_{-k_{1}}^{\dagger}a_{k_{2}}a_{k_{3}}a_{-k_{4}}^{\dagger}\rangle_{0}}\right)
 \Big[\ensuremath{\left(\delta_{k_{1},k_{2}}+\delta_{k_{1},-k_{2}}\right)}\ensuremath{\left(\delta_{k_{3},k_{4}}+\delta_{k_{3},-k_{4}}\right)}+\ensuremath{\left(\delta_{k_{1},k_{3}}+\delta_{k_{1},-k_{3}}\right)}\ensuremath{\left(\delta_{k_{2},k_{4}}+\delta_{k_{2},-k_{4}}\right)}\Big]\Bigg\}.
 \label{eq:C4}
\end{align}

Using the canonical commutation relations we can bring each of the
operator products above, in the order $a^{\dagger}a\, a^{\dagger}a$
so that we can focus only on this term and recover the others at the
end by considering permutations of the indices $1,2,3,4$. This re-ordering
procedure introduces additional lower-order terms of the form $a^{\dagger}a$
which will be taken into account below. Due to the translational invariance
of the initial state that enforces $\sum_{i=1}^{4}k_{i}=0$ for the
initial expectation values, only one out of the four pairing combinations
for each term survives in the thermodynamic limit. Therefore we have
\begin{align*}
 & \frac{1}{L^{2}}\sum_{k_{1},k_{2},k_{3},k_{4}}\frac{1}{4\sqrt{\prod_{i=1}^{4}\omega_{k_{i}}}}\; e^{i\sum_{i=1}^{4}k_{i}x_{i}}\langle a_{-k_{1}}^{\dagger}a_{k_{2}}a_{-k_{3}}^{\dagger}a_{k_{4}}\rangle_{0}\ensuremath{\left(\delta_{k_{1},k_{2}}+\delta_{k_{1},-k_{2}}\right)}\ensuremath{\left(\delta_{k_{3},k_{4}}+\delta_{k_{3},-k_{4}}\right)}=\\
 & \frac{1}{L^{2}}\sum_{k_{1},k_{2},k_{3},k_{4}}\frac{1}{4\sqrt{\prod_{i=1}^{4}\omega_{k_{i}}}}\; e^{i\sum_{i=1}^{4}k_{i}x_{i}}\langle a_{-k_{1}}^{\dagger}a_{k_{2}}a_{-k_{3}}^{\dagger}a_{k_{4}}\rangle_{0}\ensuremath{\delta_{k_{1},-k_{2}}}\ensuremath{\delta_{k_{3},-k_{4}}}=\\
& \frac{1}{L^{2}}\sum_{k,p}\frac{1}{4\omega_{k}\omega_{p}}\; e^{ik(x_{2}-x_{1})+ip(x_{4}-x_{3})}\langle n_{k}n_{p}\rangle_{0}\,,
\end{align*}
and
\begin{align*}
 & \frac{1}{L^{2}}\sum_{k_{1},k_{2},k_{3},k_{4}}\frac{1}{4\sqrt{\prod_{i=1}^{4}\omega_{k_{i}}}}\; e^{i\sum_{i=1}^{4}k_{i}x_{i}}\langle a_{-k_{1}}^{\dagger}a_{k_{2}}a_{-k_{3}}^{\dagger}a_{k_{4}}\rangle_{0}\ensuremath{\left(\delta_{k_{1},k_{4}}+\delta_{k_{1},-k_{4}}\right)}\ensuremath{\left(\delta_{k_{2},k_{3}}+\delta_{k_{2},-k_{3}}\right)}=\\
 & \frac{1}{L^{2}}\sum_{k_{1},k_{2},k_{3},k_{4}}\frac{1}{4\sqrt{\prod_{i=1}^{4}\omega_{k_{i}}}}\; e^{i\sum_{i=1}^{4}k_{i}x_{i}}\langle a_{-k_{1}}^{\dagger}a_{k_{2}}a_{-k_{3}}^{\dagger}a_{k_{4}}\rangle_{0}\ensuremath{\delta_{k_{1},-k_{4}}}\ensuremath{\delta_{k_{3},-k_{2}}}=\\
 & \frac{1}{L^{2}}\sum_{k_{1},k_{2},k_{3},k_{4}}\frac{1}{4\sqrt{\prod_{i=1}^{4}\omega_{k_{i}}}}\; e^{i\sum_{i=1}^{4}k_{i}x_{i}}\left(\langle a_{-k_{1}}^{\dagger}a_{k_{4}}a_{-k_{3}}^{\dagger}a_{k_{2}}\rangle_{0}+\langle a_{-k_{1}}^{\dagger}a_{k_{4}}\rangle_{0}\right)\ensuremath{\delta_{k_{1},-k_{4}}}\ensuremath{\delta_{k_{3},-k_{2}}}=\\
 & \frac{1}{L^{2}}\sum_{k,p}\frac{1}{4\omega_{k}\omega_{p}}\; e^{ik(x_{4}-x_{1})+ip(x_{2}-x_{3})}\left(\langle n_{k}n_{p}\rangle_{0}+\langle n_{k}\rangle_{0}\right).
\end{align*}
Finite volume corrections have been omitted in the above expressions.
Taking into account all other ($2\times6=12$) terms in (\ref{eq:C4}),
which can be obtained by re-ordering and permutation of indices, and
symmetrising the terms containing $\langle n_{k}n_{p}\rangle_{0}$
under the exchange $k\leftrightarrow p$ (since $\langle n_{k}n_{p}\rangle_{0}=\langle n_{p}n_{k}\rangle_{0}$),
we finally find 
\begin{align}
  C_{\infty}^{(4)}(x_{1},x_{2},x_{3},x_{4})=&
 \sum_{{\text{all perm.s}\atop \text{of 1,2,3,4}}}\Bigg[\frac{1}{2}\frac{1}{L^{2}}\sum_{k,p}\frac{1}{4\omega_{k}\omega_{p}}e^{ik(x_{2}-x_{1})+ip(x_{4}-x_{3})}\langle n_{k}n_{p}\rangle_{0}+\nonumber \\
 & +\frac{1}{2}\left(\frac{1}{L}\sum_{k}\frac{1}{2\omega_{k}}e^{ik(x_{2}-x_{1})}\langle n_{k}\rangle_{0}\right)\left(\frac{1}{L}\sum_{p}\frac{1}{2\omega_{p}}e^{ip(x_{4}-x_{3})}\right)\Bigg]+\nonumber \\
 & +\sum_{{\text{perm.s}\atop 2\leftrightarrow3,2\leftrightarrow4}}\left(\frac{1}{L}\sum_{k}\frac{1}{2\omega_{k}}e^{ik(x_{2}-x_{1})}\right)\left(\frac{1}{L}\sum_{p}\frac{1}{2\omega_{p}}e^{ip(x_{4}-x_{3})}\right)\nonumber \\
 & =\frac{1}{2}\frac{1}{L^{2}}\sum_{k,p}\frac{1}{4\omega_{k}\omega_{p}}
 \Bigg(\sum_{{\text{all perm.s}\atop \text{of 1,2,3,4}}}e^{ik(x_{2}-x_{1})+ip(x_{4}-x_{3})}\Bigg)
 \left(\langle n_{k}n_{p}\rangle_{0}+\langle n_{k}\rangle_{0}+\frac{1}{4}\right),
 \label{eq:lt1}
\end{align}
or, in the thermodynamic limit 
\begin{align}
 & C_{\infty}^{(4)}(x_{1},x_{2},x_{3},x_{4})=\frac{1}{2}\int\frac{dk\, dp}{(2\pi)^{2}}\frac{1}{4\omega_{k}\omega_{p}}F(k,p;x_{1},x_{2},x_{3},x_{4})\left(\langle n_{k}n_{p}\rangle_{0}+\langle n_{k}\rangle_{0}+\frac{1}{4}\right),\label{eq:lt2}
\end{align}
where we have set 
\be
F(k,p;x_{1},x_{2},x_{3},x_{4})\equiv\sum_{{\text{all perm.s}\atop \text{of 1,2,3,4}}}e^{ik(x_{2}-x_{1})+ip(x_{4}-x_{3})}.
\ee
The function $F(k,p;x_{1},x_{2},x_{3},x_{4})$ is even in both $k$
and $p$ and symmetric under their interchange. Therefore we can replace
the expression $\left(\langle n_{k}n_{p}\rangle_{0}+\langle n_{k}\rangle_{0}+\frac{1}{4}\right)$
in the sum by 
\begin{align}
 & \frac{1}{4}\left\langle \left(n_{k}+n_{-k}+1\right)\left(n_{p}+n_{-p}+1\right)\right\rangle _{0}=\nonumber\\
 & =\frac{1}{4}\left(\left\langle n_{k}n_{p}\right\rangle _{0}+\left\langle n_{-k}n_{p}\right\rangle _{0}+\left\langle n_{k}n_{-p}\right\rangle _{0}+\left\langle n_{-k}n_{-p}\right\rangle _{0}+\left\langle n_{k}\right\rangle _{0}+\left\langle n_{-k}\right\rangle _{0}+\left\langle n_{p}\right\rangle _{0}+\left\langle n_{-p}\right\rangle _{0}+1\right),
\end{align}
to obtain the alternative form 
\begin{align}
 & C_{\infty}^{(4)}(x_{1},x_{2},x_{3},x_{4})=\frac{1}{8}\int\frac{dk\, dp}{(2\pi)^{2}}\frac{1}{4\omega_{k}\omega_{p}}F(k,p;x_{1},x_{2},x_{3},x_{4})\left\langle \left(n_{k}+n_{-k}+1\right)\left(n_{p}+n_{-p}+1\right)\right\rangle _{0}.\label{eq:lt2-1}
\end{align}

From the above relations we see that the large time asymptotic of
the four-point function \emph{does depend} on the initial correlations $\langle n_{k}n_{p}\rangle_{0}$.
For an arbitrary initial state, the latter are generally independent
of $\langle n_{k}\rangle_{0}$ and therefore the GGE, which contains
only information about $\langle n_{k}\rangle_{0}$ and not about $\langle n_{k}n_{p}\rangle_{0}$,
would not predict correctly the large time four-point function. Indeed,
in the GGE, since it is Gaussian and therefore Wick's theorem applies,
the prediction for the four-point function is also disconnected, that
is 
\begin{align}
 & C_{\text{GGE}}^{(4)}(x_{1},x_{2},x_{3},x_{4})=C_{\text{GGE}}^{(2)}(x_{1},x_{2})C_{\text{GGE}}^{(2)}(x_{3},x_{4})+[2\leftrightarrow3]+[2\leftrightarrow4]\label{eq:G4gge}\\
 & =\int\frac{dk\, dp}{(2\pi)^{2}}\frac{1}{4\omega_{k}\omega_{p}}e^{ik(x_{1}-x_{2})+ip(x_{3}-x_{4})}\ensuremath{\left(\langle n_{-k}\rangle_{\text{GGE}}+\langle n_{k}\rangle_{\text{GGE}}+1\right)}\ensuremath{\left(\langle n_{-p}\rangle_{\text{GGE}}+\langle n_{p}\rangle_{\text{GGE}}+1\right)}
 +[2\leftrightarrow3]+[2\leftrightarrow4].\nonumber 
\end{align}

However the situation may be different for initial states prepared
by performing a quantum quench, i.e. ground states of some Hamiltonian
since we have not yet used all properties of ground state expectation
values. A fundamental such property is the \emph{cluster decomposition}
principle, which states that at large separations between two subsets
of physical operators of an $n$-point function the latter becomes
disconnected and as we will soon show, is responsible for the validity
of the conjecture that the GGE describes the stationary behaviour.

\subsubsection{Non-interacting pre-quench Hamiltonian}

Before we proceed to the general case, let us start with the special
case of a noninteracting pre-quench Hamiltonian. In this case the
initial state is Gaussian and by Wick's theorem we have that the expectation
values of products of the conserved charges factorise 
\begin{equation}
\langle n_{k}n_{p}\rangle_{0}=\langle n_{k}\rangle_{0}\langle n_{p}\rangle_{0}.\label{eq:gauss}
\end{equation}
Therefore the correlations $\langle n_{k}n_{p}\rangle_{0}$ do not
contain any more information than $\langle n_{k}\rangle_{0}$ and the
GGE is capable of describing the large time asymptotics of the four-point
function. Indeed, in view of the factorisation property (\ref{eq:gauss}),
the two expressions (\ref{eq:lt2-1}) and (\ref{eq:G4gge}) are identical.
The same is true for all higher order correlation functions too: as
long as the post-quench dispersion relation satisfies the previously
mentioned condition for equilibration, the correlation functions of
any order tend for large times to stationary values which are related
to the initial expectation values of products of occupation number
operators $\langle\prod_{i}n_{k_{i}}\rangle_{0}$. By Wick's theorem,
these equal to the products of the expectation values $\langle\prod_{i}n_{k_{i}}\rangle_{0}=\prod_{i}\langle n_{k_{i}}\rangle_{0}$.
According to the above discussion this means that the GGE predicts
correctly their values. We therefore see that \emph{for noninteracting
pre-quench Hamiltonians, the GGE is valid, provided that
equilibration occurs, and this is a direct consequence of Wick's theorem}.

\subsubsection{Interacting pre-quench Hamiltonian}

We now consider the more general case of an arbitrary interacting
pre-quench Hamiltonian. In this case the initial state is non-Gaussian
and Wick's theorem does not apply. To take advantage of the cluster
decomposition principle, we should first express the initial correlations
$\langle n_{k}n_{p}\rangle_{0}$ in terms of the fields $\phi(x)$
and $\pi(x)$ whose initial correlations are known from the ground
state properties of the pre-quench theory. From (\ref{eq:1}) we know
that $n_{k}+n_{-k}+1=\omega_{k}\tilde{\phi}_{k}\tilde{\phi}_{-k}+\tilde{\pi}_{k}\tilde{\pi}_{-k}/\omega_{k}$,
therefore (\ref{eq:lt1}) or (\ref{eq:lt2-1}) becomes 
\begin{align}
  C_{\infty}^{(4)}(x_{1},x_{2},x_{3},x_{4})=&
 \frac{1}{32}\frac{1}{L^{2}}\sum_{k,p}F(k,p;x_{1},x_{2},x_{3},x_{4})\bigg(\left\langle \tilde{\phi}_{k}\tilde{\phi}_{-k}\tilde{\phi}_{p}\tilde{\phi}_{-p}\right\rangle _{0}+\frac{1}{\omega_{k}^{2}}\left\langle \tilde{\pi}_{k}\tilde{\pi}_{-k}\tilde{\phi}_{p}\tilde{\phi}_{-p}\right\rangle _{0}\nonumber \\
 & \qquad+\frac{1}{\omega_{p}^{2}}\left\langle \tilde{\phi}_{k}\tilde{\phi}_{-k}\tilde{\pi}_{p}\tilde{\pi}_{-p}\right\rangle _{0}+\frac{1}{\omega_{k}^{2}\omega_{p}^{2}}\left\langle \tilde{\pi}_{k}\tilde{\pi}_{-k}\tilde{\pi}_{p}\tilde{\pi}_{-p}\right\rangle _{0}\bigg).
\end{align}
Notice that the large time value of the four-point correlation function
depends solely on four-point initial correlations, not on lower order
correlations of the initial state. Let us first focus on the first
term of this sum, and write it in coordinate space 
\begin{align}
 & \frac{1}{32 L^{2}}\sum_{k,p}F(k,p;x_{1},x_{2},x_{3},x_{4})\left\langle \tilde{\phi}_{k}\tilde{\phi}_{-k}\tilde{\phi}_{p}\tilde{\phi}_{-p}\right\rangle _{0}\nonumber \\
 & =\frac{1}{32 L^{4}}\int_{-L/2}^{+L/2}dx'_{1}dx'_{2}dx'_{3}dx'_{4}\sum_{k,p}e^{ik(x'_{2}-x'_{1})+ip(x'_{4}-x'_{3})}F(k,p;x_{1},x_{2},x_{3},x_{4})\left\langle \phi(x'_{1})\phi(x'_{2})\phi(x'_{3})\phi(x'_{4})\right\rangle _{0}\nonumber \\
 & =\frac{1}{32 L^{2}}\int_{-L/2}^{+L/2}dx'_{1}dx'_{2}dx'_{3}dx'_{4}\sum_{{\text{all perm.s}\atop \text{of }x_{1},x_{2},x_{3},x_{4}}}
 \hspace{-4mm} \delta(x'_{2}-x'_{1}+x{}_{2}-x{}_{1})\delta(x'_{4}-x'_{3}+x{}_{4}-x{}_{3})
 C_{0}^{(4)}\left(s,s+x_{1}-x_{2},r,r+x_{3}-x_{4}\right)\nonumber \\
 & =\frac{1}{32 L^{2}}\sum_{{\text{all perm.s}\atop \text{of }x_{1},x_{2},x_{3},x_{4}}}\int_{-L/2}^{+L/2}dsdr\, C_{0}^{(4)}\left(s,s+x_{1}-x_{2},r,r+x_{3}-x_{4}\right).\label{eq:5}
\end{align}
The last correlator can be decomposed, completely generally, in two
parts: the disconnected and the connected one. The disconnected part
corresponds to the sum of all two-point correlations between pairs
of the four points in any combination, while the connected part corresponds
to all the rest correlations that are present in the four-point function, i.e. 
\begin{align}
 & C_{0}^{(4)}\left(s,s+x_{1}-x_{2},r,r+x_{3}-x_{4}\right)=\nonumber \\
 & =C_{0}^{(2)}\left(s,s+x{}_{1}-x{}_{2}\right)C_{0}^{(2)}\left(r,r+x{}_{3}-x{}_{4}\right)
 +C_{0}^{(2)}\left(s,r\right)C_{0}^{(2)}\left(s+x{}_{1}-x{}_{2},r+x{}_{3}-x{}_{4}\right)\nonumber \\
 & \qquad+C_{0}^{(2)}\left(s,r+x{}_{3}-x{}_{4}\right)C_{0}^{(2)}\left(s+x{}_{1}-x{}_{2},r\right)
 +C_{0,\text{conn}}^{(4)}\left(s,s+x{}_{1}-x{}_{2},r,r+x{}_{3}-x{}_{4}\right)\nonumber \\& 
 =C_{0}^{(2)}(x{}_{1}-x{}_{2})C_{0}^{(2)}(x{}_{3}-x{}_{4})
 +C_{0}^{(2)}(r-s)C_{0}^{(2)}(r-s+x{}_{3}-x{}_{4}+x{}_{2}-x{}_{1})
 \nonumber \\& \qquad+C_{0}^{(2)}(r-s+x{}_{3}-x{}_{4})C_{0}^{(2)}(r-s+x{}_{2}-x{}_{1})
 +C_{0,\text{conn}}^{(4)}(0,x{}_{1}-x{}_{2},r-s,r-s+x{}_{3}-x{}_{4}),
\end{align}
where $C_{0}^{(2)}$ and $C_{0,\text{conn}}^{(4)}$ are the initial
two-point correlation function and connected four-point correlation
function respectively. For ground states of noninteracting Hamiltonians,
by application of Wick's theorem, the four-point function is exactly
equal to the disconnected part, therefore the connected one vanishes.

We now substitute the above expansion into (\ref{eq:5}). The first
term in the last sum does not depend on $r$ and $s$ and therefore
the integration over these variables results simply in an $L^{2}$
factor that cancels the overall $L^{-2}$ prefactor. The other three
terms are functions of $r-s$ and therefore from the integration over $r+s$ we obtain a single $L$ factor, while
after performing the remaining integration over $r-s$ 
these terms scale slower than the first one. 
This is because the integrands of those terms are decaying functions
of the distance $r-s$. In particular the decay of the connected term
$C_{0,\text{conn}}^{(4)}(0,x,s,s+y)$ at large $|s|$ is ensured by
the cluster decomposition property, from which we know that at large
distances between any two pairs of the four points the connected part
tends to zero.

The above observations mean that only the first term gives a finite
contribution in the thermodynamic limit $L\to\infty$, while the rest
give only finite size corrections. We can work similarly for the other
correlators $\langle \tilde{\phi}_{k}\tilde{\phi}_{-k}\tilde{\pi}_{p}\tilde{\pi}_{-p}\rangle _{0}$,
$\langle \tilde{\pi}_{k}\tilde{\pi}_{-k}\tilde{\phi}_{p}\tilde{\phi}_{-p}\rangle _{0}$
and $\langle \tilde{\pi}_{k}\tilde{\pi}_{-k}\tilde{\pi}_{p}\tilde{\pi}_{-p}\rangle _{0}$,
however there is a difference when we pass to the coordinate space
form of these correlators. For example, for the last one we have 
\begin{align}
 & \frac{1}{32}\frac{1}{L^{2}}\sum_{k,p}F(k,p;x_{1},x_{2},x_{3},x_{4})\frac{1}{\omega_{k}^{2}\omega_{p}^{2}}\left\langle \tilde{\pi}_{k}\tilde{\pi}_{-k}\tilde{\pi}_{p}\tilde{\pi}_{-p}\right\rangle _{0}\nonumber \\
 & =\frac{1}{32}\frac{1}{L^{4}}\int_{-L/2}^{+L/2}dx'_{1}dx'_{2}dx'_{3}dx'_{4}\sum_{k,p}e^{ik(x'_{2}-x'_{1})+ip(x'_{4}-x'_{3})}F(k,p;x_{1},x_{2},x_{3},x_{4})
 \frac{1}{\omega_{k}^{2}\omega_{p}^{2}}\left\langle \pi(x'_{1})\pi(x'_{2})\pi(x'_{3})\pi(x'_{4})\right\rangle _{0}\nonumber \\
 & =\frac{1}{32}\frac{1}{L^{2}}\int_{-L/2}^{+L/2}dsdrds'dr'\sum_{{\text{all perm.s}\atop \text{of }x_{1},x_{2},x_{3},x_{4}}}H(s')H(r')
 \left\langle \pi(s)\pi(s'+s+x_{1}-x_{2})\pi(r)\pi(r'+r+x_{3}-x_{4})\right\rangle _{0},\label{eq:6}
\end{align}
where the function $H(x)$ has been defined in (\ref{eq:H}). This
function is not a Dirac $\delta$-function as it was in the case of
$\langle \tilde{\phi}_{k}\tilde{\phi}_{-k}\tilde{\phi}_{p}\tilde{\phi}_{-p}\rangle _{0}$,
therefore we cannot reduce the number of integrals from
four to two as before. However, since the post-quench dispersion relation
is already assumed to be gapped, from (\ref{eq:H}) we can see that
$H(x)$ should typically be a function that decays exponentially with
the distance over a scale $m^{-1}$, where $m$ is the gap. Therefore
integration over the coordinate variables $s'$ and $r'$ is restricted
within a range of the order $m^{-1}$. This means that for the $L\to\infty$
behaviour of (\ref{eq:6}) one can make observations analogous to
those for (\ref{eq:5}). First we expand $\left\langle \pi(s)\pi(s'+x_{1}-x_{2})\pi(r)\pi(r'+x_{3}-x_{4})\right\rangle _{0}$
as 
\begin{align}
 & D_{0}^{(4)}(s,s'+s+x_{1}-x_{2},r,r'+r+x_{3}-x_{4})= \\
 & =D_{0}^{(2)}(s,s'+s+x{}_{1}-x{}_{2})D_{0}^{(2)}(r,r'+r+x{}_{3}-x{}_{4})
 +D_{0}^{(2)}(s,r)D_{0}^{(2)}(s'+s+x{}_{1}-x{}_{2},r'+r+x{}_{3}-x{}_{4})\nonumber \\
 & \qquad+D_{0}^{(2)}\left(s,r'+x{}_{3}-x{}_{4}\right)D_{0}^{(2)}\left(s'+s+x{}_{1}-x{}_{2},r\right)
 +D_{0,\text{conn}}^{(4)}\left(s,s'+x{}_{1}-x{}_{2},r,r'+x{}_{3}-x{}_{4}\right)=\nonumber \\
 & =D_{0}^{(2)}(x{}_{1}-x{}_{2}+s')D_{0}^{(2)}(x{}_{3}-x{}_{4}+r')
 +D_{0}^{(2)}(r-s)D_{0}^{(2)}(r'-s'+r-s+x{}_{3}-x{}_{4}+x{}_{2}-x{}_{1})\nonumber \\
 & \qquad+D_{0}^{(2)}(r'+x{}_{3}-x{}_{4})D_{0}^{(2)}(r-r'-s'-s+x{}_{2}-x{}_{1})
 +D_{0,\text{conn}}^{(4)}(0,x{}_{1}-x{}_{2}+s'-s,r-s,r'-s+x{}_{3}-x{}_{4}).\nonumber
\end{align}
Substituting into (\ref{eq:6}) we see that, for the same reasons
as before, it is only the first term in the above expansion that gives
a finite contribution in the thermodynamic limit. Therefore we have
\begin{align}
 & \frac{1}{32}\frac{1}{L^{2}}\sum_{k,p}F(k,p;x_{1},x_{2},x_{3},x_{4})\frac{1}{\omega_{k}^{2}\omega_{p}^{2}}\left\langle \tilde{\pi}_{k}\tilde{\pi}_{-k}\tilde{\pi}_{p}\tilde{\pi}_{-p}\right\rangle _{0}\nonumber \\
 & =\frac{1}{32}\sum_{{\text{all perm.s}\atop \text{of }x_{1},x_{2},x_{3},x_{4}}}\int ds'H(s')D_{0}^{(2)}(x{}_{1}-x{}_{2}+s')
 \int dr'H(r')D_{0}^{(2)}(x{}_{3}-x{}_{4}+r').\label{eq:6-1}
\end{align}
Summing all terms together we find that 
\begin{align}
C_{\infty}^{(4)}(x_{1},x_{2},x_{3},x_{4}) & =\frac{1}{32}\sum_{{\text{all perm.s}\atop \text{of }x_{1},x_{2},x_{3},x_{4}}}\left(C_{0}^{(2)}(x{}_{1}-x{}_{2})+\int ds'H(s')D_{0}^{(2)}(x{}_{1}-x{}_{2}+s')\right)\nonumber \\
 & \times\left(C_{0}^{(2)}(x{}_{3}-x{}_{4})+\int dr'H(r')D_{0}^{(2)}(x{}_{3}-x{}_{4}+r')\right),\label{eq:C4a}
\end{align}
and using our result (\ref{eq:Cinfty}) for the two-point correlation
function 
\begin{align}
C_{\infty}^{(4)}(x_{1},x_{2},x_{3},x_{4}) & =\frac{1}{8}\sum_{{\text{all perm.s}\atop \text{of }x_{1},x_{2},x_{3},x_{4}}}C_{\infty}^{(2)}(x{}_{1}-x{}_{2})C_{\infty}^{(2)}(x{}_{3}-x{}_{4})
=C_{\infty}^{(2)}(x{}_{1}-x{}_{2})C_{\infty}^{(2)}(x{}_{3}-x{}_{4})+[2\leftrightarrow3]+[2\leftrightarrow4].\label{eq:C4b}
\end{align}
Comparing with the GGE expression (\ref{eq:G4gge}), we finally conclude
that the stationary four-point function $C_{\infty}^{(4)}(x_{1},x_{2},x_{3},x_{4})$
is equal to GGE prediction, also for this more general choice of initial
state.

\subsection{Higher order correlations}

We can now outline the generalisation of the above proof to the case
of the $2n$-point function. The stationary value of the latter $C_{\infty}^{(2n)}(\left\{ x_{i}\right\} )$
would turn out to depend on $\langle\prod_{j=1}^{n}(n_{k_{j}}+n_{-k_{j}}+1)\rangle_{0}$
which in turn depends on $\langle\prod_{j=1}^{n}\tilde{\phi}_{k_{j}}^{(\sigma_{j})}\tilde{\phi}_{-k_{j}}^{(\sigma_{j})}\rangle_{0}$
where $\sigma_{j}=0,1$ and $\tilde{\phi}_{k_{j}}^{(0)}\equiv\tilde{\phi}_{k_{j}}$,
$\tilde{\phi}_{k_{j}}^{(1)}\equiv\tilde{\pi}_{k_{j}}=\dot{\tilde{\phi}}{}_{k_{j}}$.
The initial correlator $\langle\prod_{j=1}^{n}\tilde{\phi}_{k_{j}}\tilde{\phi}_{-k_{j}}\rangle_{0}$
can be expressed in terms of $\langle\prod_{j=1}^{n}\tilde{\phi}(s_{j}+x_{2j-1})\tilde{\phi}(s_{j}+x_{2j})\rangle_{0}$
which in the thermodynamic limit $L\to\infty$, due to the cluster
decomposition property of the initial state, tends to the maximally-disconnected
form $\prod_{j=1}^{n}\langle\tilde{\phi}(x_{2j-1})\tilde{\phi}(x_{2j})\rangle_{0}$.
Working similarly for the other initial correlators, we can reconstruct the GGE prediction.

\section{Non-relativistic free bosons}\label{LL}

In this second part of our study, we will check the validity of the GGE
for a quantum quench of the interaction $c$ in the Lieb-Liniger model
from arbitrary initial $c>0$ to $c=0$, a quench already studied in great detail in \cite{Mossel}. 
We first show that, evolving under this free Hamiltonian, the verification of the conjecture of validity of the GGE 
for the $g_{1}$ function (the equal time two-point function) is trivial
and tautological. This is not true for the $g_{2}$ function (the non-local pair correlation function).

The initial state $|\Omega\rangle$ is the ground state of the pre-quench
Hamiltonian
\be
H_{0}=\int_{0}^{L}dx\,\left(\partial_{x}\Psi^{\dagger}(x)\partial_{x}\Psi(x)+c\Psi^{\dagger}(x)\Psi^{\dagger}(x)\Psi(x)\Psi(x)\right),
\ee
while the evolution is described by the free boson Hamiltonian 
\be
H=\int_{0}^{L}dx\,\partial_{x}\Psi^{\dagger}(x)\partial_{x}\Psi(x)=\sum_{k=-\infty}^{+\infty}k^{2}\Psi_{k}^{\dagger}\Psi_{k},
\ee
where
\be
\Psi_{k}=\int_{0}^{L}\frac{dx}{\sqrt{L}}e^{-ikx}\Psi(x), \qquad 
\Psi(x)=\frac{1}{\sqrt{L}}\sum_{k=-\infty}^{+\infty}e^{+ikx}\Psi_{k}.
\ee
The boson mass is set to $m={1}/{2}$, the system size is $L$
and periodic boundary conditions have been assumed, so that $k=2\pi n/L$
with $n$ integer. The time evolution of the mode operators is
\be
\Psi_{k}(t)=e^{iHt}\Psi_{k}(0)e^{-iHt}=\Psi_{k}(0)e^{-ik^{2}t}.
\ee

\subsection{The $g_{1}$ function}

The $g_{1}$ function is
\be
g_{1}(x;t)=\langle\Omega|\Psi^{\dagger}(0;t)\Psi(x;t)|\Omega\rangle=\frac{1}{L}\sum_{k=-\infty}^{+\infty}e^{ikx}\langle\Omega|\Psi_{k}^{\dagger}(0)\Psi_{k}(0)|\Omega\rangle=\frac{1}{L}\sum_{k=-\infty}^{+\infty}e^{ikx}\langle n_{k}\rangle_{0},
\ee
where $n_{k} \equiv \Psi_k^\dagger \Psi_k$ are the mode occupation number operators, i.e. the conserved charges. 
We see that $g_{1}(x;t)$ is actually time independent and automatically described by the
GGE, since it is a linear combination of the values of 
$n_{k}$ in the initial state which,
by definition of the GGE, are equal to their GGE values. Explicitly
\begin{align}
g_{1,\text{GGE}}(x) & =\langle\Psi^{\dagger}(0)\Psi(x)\rangle_{\text{GGE}}
=\frac{1}{L}\sum_{k=-\infty}^{+\infty}e^{ikx}\text{\,\ensuremath{\frac{\text{Tr}\{\Psi_{k}^{\dagger}\Psi_{k}e^{-\sum_{k'}\lambda_{k'}n_{k'}}\}}{\text{Tr}\{e^{-\sum_{k'}\lambda_{k'}n_{k'}}\}}}}
=\frac{1}{L}\sum_{k=-\infty}^{+\infty}e^{ikx}\langle n_{k}\rangle_{\text{GGE}}.
\end{align}
Therefore, since $\langle n_{k}\rangle_{\text{GGE}} = \langle n_{k}\rangle_{0}$, we have
\begin{equation}
g_{1,\text{GGE}}(x)=g_{1}(x;t\to\infty)=g_{1}(x;0).\label{eq:g1}
\end{equation}
This result was previously obtained in \cite{Mossel}.

\subsection{The $g_{2}$ function}

The $g_{2}$ function is 
\begin{align}
 & g_{2}(x;t)=\langle\Omega|\Psi^{\dagger}(x;t)\Psi^{\dagger}(0;t)\Psi(x;t)\Psi(0;t)|\Omega\rangle\nonumber \\
 & =L^{-2}\sum_{k_{1},k_{2},k_{3}}e^{i\left(k_{1}-k_{3}\right)x+i\left[k_{1}^{2}+k_{2}^{2}-k_{3}^{2}-\left(k_{1}+k_{2}-k_{3}\right)^{2}\right]t}\langle\Omega|\Psi_{k_{1}}^{\dagger}(0)\Psi_{k_{2}}^{\dagger}(0)\Psi_{k_{3}}(0)\Psi_{k_{1}+k_{2}-k_{3}}(0)|\Omega\rangle\nonumber \\
 & =L^{-2}\sum_{k_{1},k_{2},k_{3}}e^{ik_{13}\left(x-2k_{23}t\right)}\langle\Omega|\Psi_{k_{1}}^{\dagger}(0)\Psi_{k_{2}}^{\dagger}(0)\Psi_{k_{3}}(0)\Psi_{k_{1}+k_{2}-k_{3}}(0)|\Omega\rangle\nonumber \\
 & =L^{-4}\sum_{k_{1},k_{2},k_{3}}e^{ik_{13}\left(x-2k_{23}t\right)}\int_{0}^{L}dx_{1}dx_{2}dx_{3}dx_{4}\, e^{-ik_{1}x_{1}-ik_{2}x_{2}+ik_{3}x_{3}+ik_{4}x_{4}}\times\nonumber \\
 & \qquad\langle\Omega|\Psi^{\dagger}(x_{1};0)\Psi^{\dagger}(x_{2};0)\Psi(x_{3};0)\Psi(x_{4};0)|\Omega\rangle\\
 & =L^{-3}\sum_{k_{13},k_{23},k_{3}}e^{ik_{13}\left(x-2k_{23}t\right)}\int_{0}^{L}dx_{1}dx_{2}dx_{3}\, e^{-ik_{13}x_{1}-ik_{23}x_{2}+ik_{3}\left(x_{3}-x_{1}-x_{2}\right)}\times\nonumber \\
 & \qquad\langle\Omega|\Psi^{\dagger}(x_{1};0)\Psi^{\dagger}(x_{2};0)\Psi(x_{3};0)\Psi(0;0)|\Omega\rangle\\
 & =L^{-2}\sum_{q,k}e^{iq\left(x-2kt\right)}\int_{0}^{L}dx_{1}dx_{2}\, e^{-iqx_{1}-ikx_{2}}\langle\Omega|\Psi^{\dagger}(x_{1};0)\Psi^{\dagger}(x_{2};0)\Psi(x_{1}+x_{2};0)\Psi(0;0)|\Omega\rangle\nonumber \\
 & =L^{-1}\sum_{k}\int_{0}^{L}dz\, e^{-ikz}\langle\Omega|\Psi^{\dagger}(x-2kt;0)\Psi^{\dagger}(z;0)\Psi(x-2kt+z;0)\Psi(0;0)|\Omega\rangle.\label{eq:g2t}
\end{align}
where $k_{ij}\equiv k_i - k_j$ and we made use of the translational invariance of the system. 
The last expression has been derived also in \cite{Mossel}. If we
decompose the initial four-point correlation function of the last
line into disconnected and connected pieces, we have 
\begin{align*}
 & \langle\Omega|\Psi^{\dagger}(x-2kt;0)\Psi^{\dagger}(z;0)\Psi(x-2kt+z;0)\Psi(0;0)|\Omega\rangle\\
 & =\langle\Omega|\Psi^{\dagger}(x-2kt;0)\Psi(x-2kt+z;0)|\Omega\rangle\langle\Omega|\Psi^{\dagger}(z;0)\Psi(0;0)|\Omega\rangle
 \\& 
 +\langle\Omega|\Psi^{\dagger}(x-2kt;0)\Psi(0;0)|\Omega\rangle\langle\Omega|\Psi^{\dagger}(z;0)\Psi(x-2kt+z;0)|\Omega\rangle\\
 & +\langle\Omega|\Psi^{\dagger}(x-2kt;0)\Psi^{\dagger}(z;0)\Psi(x-2kt+z;0)\Psi(0;0)|\Omega\rangle_{\text{conn}}\\
 & =\langle\Omega|\Psi^{\dagger}(0;0)\Psi(z;0)|\Omega\rangle\langle\Omega|\Psi^{\dagger}(z;0)\Psi(0;0)|\Omega\rangle\\
 & +\langle\Omega|\Psi^{\dagger}(x-2kt;0)\Psi(0;0)|\Omega\rangle\langle\Omega|\Psi^{\dagger}(0;0)\Psi(x-2kt;0)|\Omega\rangle\\
 & +\langle\Omega|\Psi^{\dagger}(x-2kt;0)\Psi^{\dagger}(z;0)\Psi(x-2kt+z;0)\Psi(0;0)|\Omega\rangle_{\text{conn}}\\
 & =\left|g_{1}(z;0)\right|^{2}+\left|g_{1}(x-2kt;0)\right|^{2}+G_{\text{conn}}^{(4)}(x-2kt,z,x-2kt+z,0;0),
\end{align*}
where $G_{\text{conn}}^{(4)}$ is the connected part of the four-point
function. Once again we used the fact that the initial state
is translationally invariant. We now find 
\begin{align}
g_{2}(x;t) & =\frac{1}{L}\sum_{k}\int_{0}^{L}dz\, e^{-ikz}\left(\left|g_{1}(z;0)\right|^{2}+\left|g_{1}(x-2kt;0)\right|^{2}+G_{\text{conn}}^{(4)}(x-2kt,z,x-2kt+z,0;0)\right)\nonumber \\
 & =\left(g_{1}(0;0)\right)^{2}+\left|g_{1}(x;0)\right|^{2}+\frac{1}{L}\sum_{k}\int_{0}^{L}dz\, e^{-ikz}G_{\text{conn}}^{(4)}(x-2kt,z,x-2kt+z,0;0),
\end{align}
where in the first term we performed first the $k$-summation and
then the $z$-integration, while in the second we performed them in
reverse order.

Next we take first the infinite system size limit and then the large
time limit of the last term. For $L\to\infty$ the sum becomes an
integral and the above expression becomes 
\be
\int_{-\infty}^{+\infty}\frac{dk}{2\pi}\int_{-\infty}^{+\infty}dz\, e^{-ikz}G_{\text{conn}}^{(4)}(x-2kt,z,x-2kt+z,0;0)\equiv I(x,t).
\ee
The function $G_{\text{conn}}^{(4)}(x-2kt,z,x-2kt+z,0;0)$ is regular
as $z\to0$ and $k\to x/2t$, because the initial four-point function
itself is regular when each of its coordinates tends to zero. This
is true for any initial state that is the ground state of the Lieb-Liniger model 
for some $c>0$. On the other hand, the large distance behaviour of
the connected four-point function is given by the harmonic fluid or
Luttinger liquid approximation, which states that $G_{\text{conn}}^{(4)}$
decays as a power law (along with oscillating subleading corrections).
The exponent of this power law is given in terms of the Luttinger
parameter $K$ which is related to the interaction $c$ of the Lieb-Liniger model and varies monotonically from $+\infty$ to 1 as
$c$ varies from 0 to $\infty$. 
The explicit expression for the connected four-point function is derived in the appendix \ref{app} using Conformal Field Theory methods. 
From Eq. (\ref{cft4pt}) we find that
\be
G_{\text{conn}}^{(4)}(x-2kt,z,x-2kt+z,0;0)\sim\rho_{0}^{2}\left(\left|\frac{1}{z^{2}}-\frac{1}{(x-2kt)^{2}}\right|^{1/(2K)}-\frac{1}{|z|^{1/K}}-\frac{1}{|x-2kt|^{1/K}}\right).
\ee
Replacing $G_{\text{conn}}^{(4)}$ by this asymptotic expression in
the integral does not affect its large time behaviour, since both
the exact and the asymptotic expressions are integrable at $z\to0$
and $k\to x/2t$ (because $K$ varies from $+\infty$ to 1 so that
$1/2K<1/2$) and the contribution of the region around these points
is subleading for $t\to\infty$. The large time behaviour can be easily
derived from the above scaling form. Indeed performing the change
of integration variables $k\to-k'/(2\sqrt{t})+x/(2t)$ and $z\to z'\sqrt{t}$
we have 
\be
I(x,t\to\infty)\sim\frac{1}{t^{1/(2K)}}\rho_{0}^{2}\int_{-\infty}^{+\infty}\frac{dk'}{2\pi}\int_{-\infty}^{+\infty}dz'\, e^{ik'z'/2}e^{-ixz'/(2\sqrt{t})}\left(\left|\frac{1}{z'^{2}}-\frac{1}{k'^{2}}\right|^{1/(2K)}-\frac{1}{|z'|^{1/K}}-\frac{1}{|k'|^{1/K}}\right).
\ee
For $t\to\infty$ the exponential $e^{-ixz'/(2\sqrt{t})}$ tends to
1, so that the value of $x$ is completely irrelevant in this limit.
Therefore we find 
\be
I(x,t\to\infty)\sim A(K)\rho_{0}^{2}\, t^{-1/(2K)}\to0,
\ee
where 
\be
A(K)\equiv \int_{-\infty}^{+\infty}\frac{dk'}{2\pi}\int_{-\infty}^{+\infty}dz'\, e^{ik'z'/2}\left(\left|\frac{1}{z'^{2}}-\frac{1}{k'^{2}}\right|^{1/(2K)}-\frac{1}{|z'|^{1/K}}-\frac{1}{|k'|^{1/K}}\right),
\ee
which is a convergent integral for all $1<K<\infty$. This scaling
law is in \emph{perfect agreement} with earlier numerical studies \cite{Mossel},
in which it was shown that $I(x,t\to\infty)$ decays with time as
a power law with an exponent that is a decreasing function of the
initial interaction $c$ and tends to $1/2$ as $c\to\infty$.

The above result means that the connected part vanishes for large
times and what remains is the disconnected parts 
\begin{align}
\lim_{t\to\infty}g_{2}(x;t) & =\left(g_{1}(0;0)\right)^{2}+\left|g_{1}(x;0)\right|^{2}.
\end{align}

Comparing with the prediction of the GGE in which Wick's theorem applies, that is
\begin{align}
g_{2,GGE}(x) & =\langle\Psi^{\dagger}(x)\Psi^{\dagger}(0)\Psi(x)\Psi(0)\rangle_{GGE} \nonumber \\
 & =\langle\Psi^{\dagger}(x)\Psi(x)\rangle_{GGE}\langle\Psi^{\dagger}(0)\Psi(0)\rangle_{GGE}+\langle\Psi^{\dagger}(x)\Psi(0)\rangle_{GGE}\langle\Psi^{\dagger}(0)\Psi(x)\rangle_{GGE} \nonumber \\
 & =\left(g_{1,GGE}(0)\right)^{2}+\left|g_{1,GGE}(x)\right|^{2},
\end{align}
and using (\ref{eq:g1}), we see that the two expressions are in agreement
and the GGE is correct for the $g_{2}$ correlation function.

Notice that the crucial point in the above argument was the fact that
the connected four-point function $G_{\text{conn}}^{(4)}$ decays
to zero at large distances. As in the previous problem we studied,
this is generally satisfied for any initial state prepared by performing
a quantum quench i.e. any state that is the ground state of some physical Hamiltonian,
due to the cluster decomposition principle. Using the same change
of integration variables as above, we conclude that $I(x,t)$ tends
for large times to zero in the same way as $G_{\text{conn}}^{(4)}$
decays to zero when each of its coordinates tend to infinity. Therefore
the validity of the GGE is once again a consequence of the cluster
decomposition principle.

\section{Conclusions}

We have shown that for a quantum quench problem whose evolution is governed by a noninteracting Hamiltonian
the existence of a (local) stationary state described by a GGE can be exclusively attributed to the cluster decomposition 
properties of the initial state. 
If the cluster decomposition holds for the initial state, we have analytically shown that in the thermodynamic and large
time limit the GGE describes multi-point correlation functions of the fields. Inversely when cluster 
decomposition does not hold, additional non-vanishing terms arise which do not comply with Wick's theorem that 
is tautologically valid in the GGE of noninteracting models. 
While the calculation has been performed for two specific free bosonic theories (gapped `relativistic' bosons and 
free non-relativist bosons), the line of the derivation is completely general and should be applicable to 
arbitrary local noninteracting systems.    
Notice that even though we tested the validity of the GGE only for equal-time correlations, a general theorem \cite{eef-12}
ensures that when this happens, different-time stationary correlations are also described by the GGE.

We also point out another interesting byproduct of this work. While the connected part of multi-point correlations in the initial state does not contribute to the stationary value, it however contributes to the approach to the GGE. As explained in the introduction, in order to avoid the quantum revivals, the thermodynamic limit should be taken before the large time limit. Then the extra terms (including the connected part of multi-point correlations) determine the time decaying part of observables, as in the calculation of section \ref{LL}. On the other hand, when it is already known that the system equilibrates, the stationary behaviour can be derived by first time-averaging over infinite time and then taking the thermodynamic limit. In this case the extra terms determine the finite size part of observables that decays with the system size, as in the discussion of section \ref{RB}. If specialised to finite but large systems, similar arguments can be used to understand revival properties of some observables, a topic of intense recent interest \cite{Mossel,c-14,ir-11}.

One may expect that or wonder if an analogous reasoning may be the basis of the validity of the conjecture that the GGE describes the stationary behaviour after a quantum quench also for the general case of a \emph{genuinely interacting} integrable post-quench Hamiltonian. Let us think about this scenario in more detail. It is true that even in this more general case, the system can be decomposed in terms of momentum (or rapidity) modes, whose occupation number operators are conserved and are linear combinations of the local conserved charges. The creation and annihilation operators of these modes evolve in time exactly as in the noninteracting case and satisfy generalised canonical commutation relations, which involve the two-particle scattering matrix of the model (in the context of relativistic Integrable Field Theory, this is known as the Zamolodchikov-Faddeev algebra \cite{sfm-12}). Using these generalised canonical commutation relations, it is possible to derive a generalised version of Wick's theorem, reducing all higher order correlations of the creation/annihilation operators to their lowest (two-particle) correlations. In the context of quantum quenches, this allows the possibility to express higher order correlations in terms of solely the initial values of the charges, as in the noninteracting case. However, unlike the noninteracting case, the local physical observables (fields or vertex operators) are given in terms of the mode creation/annihilation operators through complicated nonlinear expressions (typically series expansions involving all of their Form Factors, i.e. the matrix elements of the observables in the energy eigenstates) and therefore at this point, it is technically difficult to make a connection with the above reasoning and take advantage of the cluster decomposition principle for the initial field correlations.

\section*{Acknowledgments} We are grateful to John Cardy and Mario Collura for helpful discussions.  
This work was supported by  the ERC under  Starting Grant 279391 EDEQS (PC and SS).

\appendix

\section{The four-point function in the Luttinger model}\label{app}

In this appendix, we will calculate the connected part of the four-point
function of the Luttinger model, that gives an effective description
of the Lieb-Liniger model at large distances. The calculation is based
on the Conformal Field Theory method (for details and notation, see
\cite{Caz}). The general four-point function is

\begin{align*}
 & \langle\Psi^{\dagger}(x_{1})\Psi^{\dagger}(x_{2})\Psi(x_{3})\Psi(x_{4})\rangle=\\
 & =\rho_{0}^{2}\sum_{m_{1},m_{2},m_{3}=-\infty}^{+\infty}e^{2\pi i\rho_{0}[m_{1}x_{1}+m_{2}x_{2}-m_{3}x_{3}-(m_{1}+m_{2}-m_{3})x_{4}]}
 \langle A_{2m_{1},-1}(x_{1})A_{2m_{2},-1}(x_{2})A_{-2m_{3},+1}(x_{3})A_{-2(m_{3}-m_{1}-m_{2}),+1}(x_{4})\rangle\\
 & =\rho_{0}^{2}\langle A_{0,-1}(x_{1})A_{0,-1}(x_{2})A_{0,+1}(x_{3})A_{0,+1}(x_{4})\rangle+\cdots
\end{align*}
where $A_{m,n}=e^{im\theta}e^{in\phi}=e^{i\beta(m,-n)\phi_{L}}e^{i\beta(m,n)\phi_{R}}$
are the vertex operators with $\beta(m,n)=m\sqrt{K}/2+n/(2\sqrt{K})$, $K$ is the Luttinger parameter (related to the interaction
$c$ of the Lieb-Liniger model), and $\rho_0$ is the boson density. The dots denote oscillating terms
that decay with the distance faster than the term we kept and are
therefore negligible. The leading term is
\begin{align}
 & \langle A_{0,-1}(x_{1})A_{0,-1}(x_{2})A_{0,+1}(x_{3})A_{0,+1}(x_{4})\rangle\nonumber \\
 & =\left(\frac{2\pi}{L}\right)^{4q^{2}}\left|z_{1}z_{2}z_{3}z_{4}\right|^{q^{2}}\langle V_{-q}(z_{1})V_{-q}(z_{2})V_{q}(z_{3})V_{q}(z_{4})\rangle\langle\bar{V}_{q}(\bar{z}_{1})\bar{V}_{q}(\bar{z}_{2})\bar{V}_{-q}(\bar{z}_{3})\bar{V}_{-q}(\bar{z}_{4})\rangle\nonumber\\
 & =\left(\frac{2\pi}{L}\right)^{4q^{2}}\left|z_{1}z_{2}z_{3}z_{4}\right|^{q^{2}}\left(\frac{z_{12}z_{34}}{z_{13}z_{14}z_{23}z_{24}}\right)^{q^{2}}\left(\frac{\bar{z}_{12}\bar{z}_{34}}{\bar{z}_{13}\bar{z}_{14}\bar{z}_{23}\bar{z}_{24}}\right)^{q^{2}}\nonumber\\
 & =\left(\frac{2\pi}{L}\right)^{1/K}\left|z_{1}z_{2}z_{3}z_{4}\left(\frac{z_{12}z_{34}}{z_{13}z_{14}z_{23}z_{24}}\right)^{2}\right|^{1/(4K)}
 \nonumber\\
 & =\left(\frac{d(x_{1}-x_{2}|L)d(x_{3}-x_{4}|L)}{d(x_{1}-x_{3}|L)d(x_{1}-x_{4}|L)d(x_{2}-x_{3}|L)d(x_{2}-x_{4}|L)}\right)^{1/(2K)},
\end{align}
where $V_{q}(z)=\left.:e^{iq\phi_{L}(z)}:\right.$, $\bar{V}_{q}(\bar{z})=\left.:e^{iq\phi_{R}(\bar{z})}:\right.$
are the normal-ordered vertex operators, the expectation values are
evaluated in a cylindrical geometry (due to the periodic boundary
conditions) i.e. $z_{i}=e^{2\pi w_{i}/L}$ with $w_{i}=v\tau_{i}+ix_{i}$
(in our case the imaginary times are zero, $\tau_{i}=0$) and $q=1/(2\sqrt{K})$.
The function $d(x|L)=|\sin(\pi x/L)|L/\pi$ is the cord function and
in the infinite size limit $L\to\infty$ becomes $d(x|L)\to|x|$.
In this limit we therefore have 
\begin{align}
 & \langle\Psi^{\dagger}(x_{1})\Psi^{\dagger}(x_{2})\Psi(x_{3})\Psi(x_{4})\rangle=\rho_{0}^{2}\left|\frac{(x_{1}-x_{2})(x_{3}-x_{4})}{(x_{1}-x_{3})(x_{1}-x_{4})(x_{2}-x_{3})(x_{2}-x_{4})}\right|^{1/(2K)}+\cdots
\end{align}
Finally, the connected part of the above is 
\begin{align}
 & G_{\text{conn}}^{(4)}(x_{1},x_{2},x_{3},x_{4};0)=\nonumber \\
 & =\langle\Psi^{\dagger}(x_{1})\Psi^{\dagger}(x_{2})\Psi(x_{3})\Psi(x_{4})\rangle-\langle\Psi^{\dagger}(x_{1})\Psi(x_{3})\rangle\langle\Psi^{\dagger}(x_{2})\Psi(x_{4})\rangle-\langle\Psi^{\dagger}(x_{1})\Psi(x_{4})\rangle\langle\Psi^{\dagger}(x_{2})\Psi(x_{3})\rangle\nonumber \\
 & =\rho_{0}^{2}\Bigg(\left|\frac{(x_{1}-x_{2})(x_{3}-x_{4})}{(x_{1}-x_{3})(x_{1}-x_{4})(x_{2}-x_{3})(x_{2}-x_{4})}\right|^{1/(2K)}-\nonumber \\
 & \qquad-\frac{1}{|(x_{1}-x_{3})(x_{2}-x_{4})|^{1/(2K)}}-\frac{1}{|(x_{1}-x_{4})(x_{2}-x_{3})|^{1/(2K)}}\Bigg)+\cdots \label{cft4pt}
\end{align}

\end{document}